\documentclass[aps,prb,twocolumn,showpacs]{revtex4}
 \usepackage{amsmath}
\usepackage{graphicx}
\usepackage{epsfig}
\newcommand{\beq}{\begin{equation}}
\newcommand{\eeq}{\end{equation}}
\newcommand{\bea}{\begin{eqnarray}}
\newcommand{\eea}{\end{eqnarray}}
\newcommand{\pdag}{{\phantom{\dagger}}}

\begin{document}
\bibliographystyle{apsrev}
 
\title{ Multiple crossovers in interacting quantum wires }
\author{M.\ Kindermann and P.\ W.\ Brouwer}
\affiliation{ Laboratory of Atomic and Solid State Physics, Cornell University, Ithaca, New York 14853-2501  }

\date{\today}
\begin{abstract}
We study tunneling of electrons into and between interacting wires in the spin-incoherent regime subject to a magnetic field. The tunneling currents follow power laws of the applied voltage with  exponents that depend on whether the electron spins at the relevant length scales are polarized or disordered.  The crossover length (or energy) scale  is exponential in the applied field. In a finite size wire multiple crossovers can occur. 
       \end{abstract}
\pacs{73.63.Nm,71.10.Pm,71.27.+a}
\maketitle

\section{Introduction}

The Luttinger Liquid is considered to be the appropriate description
of the low-energy properties of interacting electrons in a quantum
wire.\cite{kn:haldane1981,kn:giamarchi2004}
In a Luttinger Liquid, the fundamental excitations are
collective charge and spin modes. Unlike Fermi Liquids, Luttinger 
Liquids have power law dependences for the tunneling density of
states and for the conductance through a weak link as a function of
the applied bias voltage or temperature. This and other properties of Luttinger Liquids such as the phenomenon of spin-charge separation have been observed experimentally. \cite{kn:yao1999,kn:bockrath1999,kn:auslaender2002,kn:ishii2003}

Recently, it has been realized that the electron liquid with
spin excitation energy $J$
smaller than the temperature $T$ differs qualitatively  from the 
standard Luttinger Liquid 
picture.\cite{kn:cheianov2004,kn:cheianov2004b,kn:cheianov2005,kn:matveev2004,kn:matveev2004b,kn:fiete2004,kn:fiete2005,kn:fiete2005b,kn:fiete2006,kn:kindermann2005,kn:kindermann2006,kn:kakashvili2006,kn:fiete2006b,kn:matveev2006}
This `spin-incoherent' limit can be realized in quantum wires at low
densities, when the conductor enters the Wigner crystal regime and $J$
is exponentially suppressed below the Fermi energy $E_F$ by the
Coulomb barrier between adjacent electrons in the
crystal.\cite{kn:klironomos2005,kn:fogler2005b} Recent experiments on
wires defined in heterostructures approach this
regime.\cite{kn:auslaender2002,kn:fiete2005}
The spin-incoherent limit can also be reached in ultra-thin quantum
wires, in
which the Coulomb repulsion renders the electrons effectively
impenetrable.\cite{kn:fogler2005} A theoretical model that shows spin-incoherence is the one-dimensional Hubbard model with infinite on-site
repulsion $U$.

\begin{figure} 
\includegraphics[width=6.5cm]{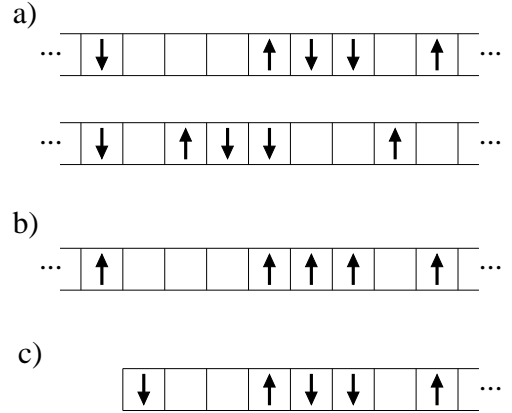}
\caption{   Hubbard chain with infinite on-site repulsion $U$ that
  allows sites to be singly occupied only. a) Moving charge carriers
  (holes in the electron occupation) shift also the spin configuration
  such that the spin state at a given site may change. The relative
  position of the spins remains fixed, however. b) The spin state at a
  given site does not fluctuate in a spin-polarized Hubbard chain,
  when the
  spin configuration is translationally invariant. c) Also the spin
  configuration at the site that is at the end of the chain cannot be
  changed by the charge dynamics. }   \label{fig1}
\end{figure}

The qualitative difference between the cases $J=0$ and $J \neq 0$
arises because for $J=0$ the velocity $v_s$ of spin excitations is
zero. This implies that the relative positions of 
the electron spins are fixed if $J=0$. Their absolute positions 
along the wire, however, are not. 
Spin can still move along the wire together with the flow of charge.
This is best visualized in the 
infinite-$U$ Hubbard model, which has an exact solution in terms of 
a static spin 
background and non-interacting spinless holes, which are the charge
carriers,\cite{kn:Bernasconi1975}
see Fig.\ \ref{fig1}. The 
same coupling between spin and
charge generically  exists also  in Luttinger Liquids with $J \neq 0$, but its impact
on the low-energy properties is overwhelmed by the effects 
of the finite spin velocity in the standard case. For spin-incoherent Luttinger Liquids, the motion of the spin background
driven by quantum fluctuations of the charge current has been shown 
to give rise to an anomalous dependence of the tunneling current 
$I$ through a probe weakly coupled 
to the quantum wire on the applied voltage $V$. This occurs because the spin
background at the point of tunneling is not static on the time scale $\simeq \hbar/e V$ of a
tunneling event. In Refs.\ \onlinecite{kn:cheianov2004,kn:fiete2004}
it was shown that \cite{footnote}
\begin{equation}
  I \propto V^{1/2}
\end{equation}
if there are no residual interactions between the charge degrees of
freedom, whereas $I \propto V$ in the $J \downarrow 0$ limit of
finite-$J$ Luttinger Liquids in the same situation.

The motion of the spin background has no effect on the tunneling
current  if a magnetic field $B$ is applied large enough to
polarize the electron spins in the quantum wire and in the probe. In
that case, the spin background is translation invariant and one
restores the 
dependence $I \propto V$ of the non-interacting case. The tunneling
current of minority electrons   remains anomalous, however, since the addition of minority electrons
to the quantum wire always breaks the translation invariance of the
wire's spin background.

A second possibility to restore the noninteracting tunneling exponent
in a spin-incoherent Luttinger Liquid is to move the tunneling probe to
the end of the wire. In that case there is no room for variation of
the positions of the spins in the wire relative to the point of tunneling. The spin of an electron
tunneling into the wire is simply
added to the end of the spin configuration 
in the wire prior to tunneling. 

In this article we ask the question how the crossover between the
different tunneling exponents takes place. For both crossovers --- as
a function of applied magnetic field or as a function of the distance
from the end of the wire --- we find that the dependence $I \propto
V^{\alpha}$ of the tunneling current on the applied voltage switches 
between two different exponents $\alpha$. For the magnetic-field
induced crossover, we find that the majority-electron current
$I \propto V^{1/2}$ for $V \ll V_{B}$, whereas $I \propto V$ for
higher bias voltages, where the crossover voltage $V_{B}$ can be 
found as the lowest voltage at which the $(E_F/eV)^{1/2}$ spins
bordering the tunneling site can still be expected to be majority spins,
\begin{equation}
  e V_{B} = E_F e^{-2 E_{\rm Z}/k_B T}.
\end{equation}
Here $E_{\rm Z}=2 \mu_B B$, where $\mu_B$ is the Bohr magneton,  is the Zeeman energy of electrons in the applied
magnetic field $B$. For the crossover that depends on the distance $x$
from the wire's end, one has $I \propto V$ for $V \ll V_{x}$,
with
\begin{equation}
  V_x = \hbar v/x,
\end{equation}
whereas $I \propto V^{1/2}$ for higher bias voltages, where
$v$ is the velocity of the collective charge modes in the
quantum wire.

Our detailed calculations are described in the remainder of this 
article. In Sec.\ \ref{sec:2} we
discuss  the bosonized description of a one-dimensional
conductor in the spin-incoherent limit. In Sec.\ \ref{sec:3} we
present a detailed calculation of the single-electron Green functions 
of the one-dimensional wire. Finally, in Sec.\ \ref{sec:4} we discuss
the consequences that the asymptotic time-dependence of the Green functions has for tunneling experiments. In our 
calculations, we take into account that there may exist additional 
forward scattering 
interactions between the charge excitations. This applies, for instance, 
to a quantum wire at low electron densities when the wire enters the
Wigner crystal regime. The charge dynamics in such a quantum wire is
described by a spinless Luttinger Liquid with interaction parameter
$g$.\cite{kn:fiete2004,kn:fiete2005b} The tunneling exponents listed
above are for the case $g=1$. For general $g$ the nature of the crossovers
is unchanged, although the estimates of the crossover voltages and
the exact tunneling exponents are modified. 

\section{Model}
\label{sec:2}

At low energies an interacting wire in the spin-incoherent regime is 
described by a spinless charge degree of freedom and a static spin 
background. \cite{kn:fiete2004} 
The spin background does not enter  the Hamiltonian. 
We describe the charge degree of freedom using boson fields $\theta$
and $\phi$, which are related to the charge density $\rho$ 
 through
\begin{equation}
  \rho = \frac{e}{\pi} \frac{\partial \theta}{\partial x}.
\end{equation} 
We consider a half-infinite quantum wire $0 < x < \infty$. For a
half-infinite wire, the boson fields have the commutation relation
\begin{equation}
  [\theta(x), \phi(x')] = -i \pi \Theta(x-x'),
\end{equation}
where $\Theta(y) = 1$ if $y > 0$ and $0$ otherwise, and satisfy the
boundary condition $\theta(x) = 0$ and $\partial \phi(x)/\partial x =
0$ at $x=0$. The Hamiltonian is
\beq \label{Hincoh}
   H = v \int_0^\infty{\frac{dx}{2\pi}
  \left[ g^{-1} (\partial_x\theta)^2+g(\partial_x\phi)^2\right]},
\eeq 
where $g$ sets the strength of the residual interactions between the
charge excitations and $v$ is the velocity of the charge carriers. One has $g < 1$ for repulsive interactions (we set $\hbar=1$).

The annihilation operator $\psi_{\sigma}(x)$ of an electron with spin
$\sigma$ at position $x$ is
expressed in terms of spinless fermions $c(x)$ and operators 
$S_{\sigma}(x)$ that remove a spin $\sigma$ from the spin background 
of the wire at position $x$ as 
\begin{equation}
  \psi_{\sigma}(x)=c^\dag(x) S_{\sigma}(x).
\end{equation}
The annihilation operator $c$ for the spinless charge carrier is
expressed in terms of the boson fields as
\begin{eqnarray} \label{field}
  c(x) &=&\frac{\eta}{\sqrt{2\pi a}}
  \\ && \mbox{} \times
  \left[e^{-i k_F x - i \theta(x) + i \phi(x)} +
  e^{i k_F x + i \theta(x) + i \phi(x)} \right].
  \nonumber
\end{eqnarray}
 Here $\eta$ is a Majorana fermion, $k_F$ is the Fermi
wavenumber of the spinless charge carriers, and $a$ the short-distance
cut-off of the bosonized theory.

In the calculations below we need the correlation functions of the
boson fields at zero temperature. Up to a (divergent) constant that
disappears from the final results, these are\cite{kn:fabrizio1995}
\begin{eqnarray}
  \label{eq:bosoncorrelator}
  -i \langle \phi(x,\tau) \theta(x,0) \rangle &=&
  \frac{i}{4} \sum_{\pm} (\mp) \ln [a + i (v \tau \pm 2 x)], 
  \nonumber \\
  -i \langle \theta(x,\tau) \phi(x,0) \rangle &=&
  \frac{i}{4} \sum_{\pm} (\pm) \ln [a + i (v \tau \pm 2 x)], \nonumber \\
  -i \langle \phi(x,\tau) \phi(x,0) \rangle &=&
  \frac{i}{2 g} \ln (a + i v \tau) \nonumber \\ && \mbox{} +
  \frac{i}{4 g} \sum_{\pm}
  \ln (a + i (v \tau \pm 2 x)),  \nonumber \\
  -i \langle \theta(x,\tau) \theta(x,0) \rangle &=&
  \frac{i g}{2} \ln (a + i v \tau) \\ && \mbox{} -
  \frac{i g}{4} \sum_{\pm}
  \ln (a + i (v \tau \pm 2 x)). \nonumber
\end{eqnarray}

The Hamiltonian (\ref{Hincoh})
can be derived microscopically as the low-energy theory of a Hubbard
model with infinite interaction parameter $U$ and additional 
long-range interactions described by a purely forward scattering 
density-density interaction. For this derivation, one describes the infinite-$U$
Hubbard model  in terms of a
static spin background and spinless holes. The latter are the 
fermions $c$ introduced above.\cite{kn:ogata1990,kn:kindermann2005}
At low
energies one may linearize the spectrum of the holes, writing the
annihilation operator of a hole as $c(x) = 
e^{ik_F x} c_{ L}(x)+e^{-ik_F x} 
c_{ R}(x)$. 
Including the long-range interactions between the hole
densities $\rho_{ L}(x) = c^\dag_{ L}(x) c^\pdag_{
  L}(x)$ and $\rho_{R}(x) = c^\dag_{R}(x) c^\pdag_{
  R}(x)$, we then find a Hamiltonian of the form
\bea \label{Hmic}
  H &=& \int_0^{\infty} dx
  \left\{-iv_F( c^\dag_{ L}\partial_x
  c^\pdag_{ L}- c^\dag_{ R}\partial_x c^\pdag_{ R})
  \right.
  \\ && \left. \mbox{} +
  \frac{g_0}{2} \left[ \rho_{ L}^2(x)+\rho_{
      R}^2(x)\right] +  \tilde{g}_0 \rho_{ L} (x)\rho_{
    R}(x) \right\}. \nonumber 
\eea
This Hamiltonian is brought into the form of Eq.\ (\ref{Hincoh}) by
bosonization of the operators $c_{L}$ and $c_{ R}$. The
phenomenological parameters $v$ and $g$ in Eq.\ (\ref{Hincoh}) are
related to those in Eq.\ (\ref{Hmic}) by $v=v_R [(1+ g_0/2\pi v_F)^2-(\tilde{g}_0/2\pi v_F)^2]^{1/2}$ and
$g=(1+ g_0/2\pi v_F-\tilde{g}_0/2\pi v_F)^{1/2} /(1+ g_0/2\pi v_F+\tilde{g}_0/2\pi v_F)^{1/2}$. The hole
operators $c$ take the form of Eq.\ (\ref{field}) after bosonization.
 
\section{Calculation}
\label{sec:3}
 
In order to compute the tunneling current $I$ from a
tunneling probe into the wire at a distance $x$ from the wire's end,
we express $I$ in terms of the Green functions $G$ and $G_{\rm p}$ of
the wire and the probe, respectively. To lowest order in the tunneling amplitude $t$ between wire and probe one has
\begin{eqnarray}  \label{tunnel}
   I &=& 2 |t|^2 \sum_{\sigma} \int d\tau\, e^{-i e V \tau/\hbar}
  \\ && \mbox{} \times
  \left[ G^<_{ \sigma}(x,\tau)G^>_{{\rm p},\sigma}(x,-\tau)-
  G^>_{\sigma}(x,\tau)G^<_{{\rm p},\sigma}(x,-\tau)\right].
  \nonumber
\end{eqnarray}
The greater and lesser Green functions of the wire are defined as
\begin{eqnarray}
  G_{\sigma}^>(x,\tau) &=& - i \langle
  \psi^\pdag_{\sigma}(x,\tau)\psi^\dag_{\sigma}(x,0)\rangle, \nonumber \\
  G_{\sigma}^<(x,\tau) &=& i \langle
  \psi^\dag_{\sigma}(x,0)\psi^\pdag_{\sigma}(x,\tau)\rangle,
\end{eqnarray}
with similar definitions for the greater and lesser Green functions of
the probe.  

The power-law dependence of the tunneling current $I$ on 
the applied voltage $V$ at low $V$ is related to the time-dependence
of the wire Green functions at large $\tau \simeq \hbar/e V$. In terms of 
hole and spin operators we have
  \beq \label{Gup}
  G_{\sigma}^>(x,\tau) = -i\langle S^\pdag_\sigma(x,\tau)c^\dag(x,\tau)  
  c(x,0)S^\dag_\sigma(x,0)\rangle .
\eeq
The spin expectation value in Eq.\ (\ref{Gup}) is non-vanishing only
if all background spins between the spin that is at position $x$ at
time $\tau$  and the spin that is at $x'$ at time $0$ have the same
orientation $\sigma$. This occurs with probability
$p_\sigma^{|N_x(\tau)-N_{x}(0)|}$, where 
\begin{equation}  N_x(\tau)= \frac{k_F x+\theta(x,\tau)}{\pi}
\end{equation}
is the number of electrons between positions $0$ and $x$ at time $\tau$
and 
\begin{equation}
  p_\uparrow=1-p_\downarrow=\frac{1}{e^{-E_{\rm Z}/k_BT}+1}
\end{equation}
is the probability for a spin to point along the direction of the applied 
magnetic field. One then evaluates the greater Green function 
as\cite{kn:fiete2004,kn:fiete2005}
\begin{eqnarray}
  G_{\sigma}^>(x,\tau) &=& 
  -ie^{i E_{\rm Z} \sigma\tau/2 }\sum_k p_\sigma^{|k|} \int \frac{d\xi}{2\pi}\,e^{i\xi k} 
   \\ && \mbox{} \times
  \langle e^{-i \xi N_x(\tau)} c^\dag(x,\tau) c(x,0)e^{i\xi
  N_{x}(0)}\rangle .\nonumber
\end{eqnarray}
 Here, $\sigma$ equals $1$ for spin-up electrons and $-1$ for spin-down electrons. Using the correlators of Eq.\
(\ref{eq:bosoncorrelator}), we then find
\begin{widetext}
\bea
  G_{\sigma}^>(x,\tau) &\propto&
  e^{i E_{\rm Z} \sigma \tau/2 }\sum_k\, \frac{p_\sigma^{|k|} e^{-k^2/2w^2} }{2 \pi w a}
  \left\{  2 (-1)^k \left[ \frac{4 x^2 a^2}{ \left((iv\tau+a)^2+4
  x^2\right)(iv  \tau+a)^2}\right]^{1/4g} \right. \nonumber \\
  &&\left. \mbox{} + \frac{(iv\tau+a)^{g/2-1/2g}
  a^{g/2+1/2g}}{(2x)^{g/2-1/2g
  }((iv\tau+a)^2+4x^2)^{(g-1)^2/4g}}\left(\frac{e^{2i k_F
    x}}{iv\tau-2ix+a}+\frac{e^{-2i k_F
    x}}{iv\tau+2ix+a}\right)\right\}
  \label{Gdir}
\eea
\end{widetext}
with
\begin{equation}
  w^2=- \frac{g}{2 \pi^2} \ln \left[\frac{a^2}{4x^2}+
    \frac{a^2}{(iv\tau+a)^2}
  \right].
\end{equation} 
The symbol ``$\propto$'' in Eq.\ (\ref{Gdir}) indicates  dimensionless
prefactors that do not depend on  $\tau$, $x$, or $p_\sigma$.
For the Green functions needed in this article the lesser functions $G^<(x,\tau)$ 
are obtained by the replacement
\beq \label{Glesser}
  G_{\sigma}^<(x,\tau) = p_\sigma e^{i E_{\rm Z} \sigma \tau  }
  G_{\sigma}^{>}(x,\tau)^{*}|_{k_F\leftrightarrow-k_F}.
\eeq
The extra factor $p_{\sigma}$ originates from the reversed order of the
spin operators $S(x,\tau)$ and $S^{\dagger}(x,\tau)$ in the expression
for $G^{<}(x,\tau)$.

We should note that our calculation using bosonization is not exact since it
does not account for the discreteness of charge. For this reason, the
sum over $k$ was replaced by an integral in Ref.\
\onlinecite{kn:kindermann2006}. With this replacement, the Green functions
at non-coinciding points agree with the exact Green functions in the
limit $p_{\sigma} \to 1$. None of the Green functions at coinciding
points considered in this article depends on whether one has a sum or
an integral in this limit. We therefore keep the sums, which results
in improved approximations in the opposite limit $p_\sigma \to 0$.

The sum over $k$ in Eq.\ (\ref{Gdir}) is time- and
space-dependent through $w$. 
In previous treatments of the problem in Refs.\
\onlinecite{kn:fiete2004,kn:fiete2005,kn:kindermann2006} this dependence has been neglected. This is a good approximation for an unpolarized electron gas at not too large energies. As one polarizes the electron spins by applying a magnetic field, however, this additional time and space-dependence in Eq.\ (\ref{Gdir}) becomes more and more important and we will show that it leads to the expected crossover of the scaling exponent to the conventional Luttinger Liquid exponent at perfect spin-polarization.

\subsection{Crossover induced by a magnetic field}
 \label{magnetic}

\begin{figure} 
\includegraphics[width=6.5cm]{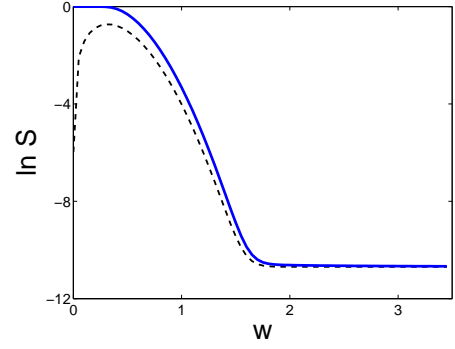}
\caption{   \label{fig5}
  Comparison between the logarithms of the sum $S$ (solid
  line) and its approximation Eq.\ (\ref{capprox}) (dashed line) at
  $E_{\rm Z}= 10 \, k_B T$.  The approximation improves as $  w$ grows
  large.}
\end{figure}

To evaluate Eq.\ (\ref{Gdir}) we need  the sum
\beq \label{sum}
 S=\sum_k\,  (-1)^k p^{|k|} e^{-k^2/2w^2} 
 \eeq
with $0\le p\le 1$. Using the identity
$$
  \sum_k f(k)=
  \int dk\, f(k)
  \sum_q e^{2i\pi  q k},
$$
we first convert the sum over $k$ into an integral,
\beq
S=2\sum_q \int_0^{\infty}dk\,  e^{-i \pi (2q+1)k+k \ln p- k^2/2w^2} .
\eeq
Although this integral can be expressed in terms of error functions,
the power-law dependence on $\tau$ can be extracted using a sequence
of approximations. For this we confine ourselves to the regime 
$\pi |w| \gg 1$, which is appropriate in the relevant case that 
the distance from the end of the wire is not microscopic ($x \gg a$) 
and the time $\tau$ is large. We first rewrite $S$ by shifting the 
contour of integration in the complex plane,
\begin{equation}
  S = S_1 + S_2,
\end{equation}
with
\begin{eqnarray*}
  S_1 &=& - 2i 
  \sum_{q} e^{-\pi^2 w^2 (2q+1)^2/2}
  \int_0^{\pi w^2(2q+1)} d\kappa\, 
  e^{\kappa^2/2 w^2}
  \nonumber \\ && \mbox{} \times e^{
  i \ln p\left(\kappa -\pi w^2(2q+1)\right)}, \nonumber \\
  S_2 &=& 2  \sum_{q} e^{w^2[\ln p-i \pi(2q+1)]^2 /2 }
  \int_{w^2 |\ln p|}^{\infty} dk\, e^{- k^2/2w^2}. \nonumber 
\end{eqnarray*}
At $\pi |w| \gg 1$ the $\kappa$ integration in $S_1$ is dominated by
$\kappa$ close to $\pi w^2 (2 q + 1)$. Linearizing the exponent around
$\kappa = \pi w^2 (2 q + 1)$, we then find
\begin{eqnarray}
  S_1 &\approx & - \sum_{q} \frac{2}{\ln p - i \pi (2 q + 1)}
  \nonumber \\ &=&
  \frac{1-p}{1+p}.
\end{eqnarray}
Each term in the $q$-summation for $S_2$ can be bounded by
$e^{-\pi^2(2 q + 1)^2 w^2/2}/|\ln p|$, from which one concludes that
$S_1$ dominates over $S_2$ unless $|\ln p| \lesssim \exp(-\pi^2
w^2/4)$. Since $\exp(-\pi^2 w^2/4) \ll 1/ w^2$, one may in our limit $\pi w \gg 1$ thus
calculate $S_2$ in the limit $\ln p \to 0$. We then find
\beq \label{capprox}
 S \approx  \frac{1-p}{1+p}+ \sqrt{2\pi}w \,e^{-w^2 \pi^2/2}
\eeq
in the regime of interest  $\pi |w| \gg 1$. The numerical comparison  shown in Fig.\ \ref{fig5}  shows that Eq.\ (\ref{capprox}) is  indeed an excellent approximation to the sum  Eq.\ (\ref{sum}). 

In order to study the magnetic-field dependence of tunneling into the
bulk of a spin-incoherent wire, we take $v\tau \ll x$ in Eq.\ (\ref{Gdir}). Using
Eq.\ (\ref{capprox}), we then find 
\begin{eqnarray}
  G_\sigma^>(\tau) &\propto&
\frac{e^{i E_{\rm Z} \sigma \tau/2  }}{a}\Bigg[  \frac{1}
  {\sqrt{2 \pi} w } \frac{1-p_\sigma}{1+p_\sigma}
  \left(\frac{a}{iv\tau+a} \right)^{1/2g}\
  \nonumber \\ && \mbox{}
    + 
    \left(\frac{a}{ iv\tau+a}\right)^{g/2+1/2 g}\Bigg].
  \label{Gh}
\end{eqnarray}
The first term has the anomalous scaling $\propto \tau^{-1/2g}$
calculated in Refs.\ \onlinecite{kn:cheianov2004,kn:fiete2004} for the
spin-incoherent Luttinger Liquid at zero magnetic field. (We neglect
the logarithmic time dependence following from the prefactor $w$ in
Eq.\ (\ref{Gh}).) The second
term has the asymptotic time-dependence characteristic of a standard
Luttinger Liquid at interaction parameter $g$. In the absence of a
magnetic field, the first term dominates. In the presence of a
magnetic field that partially polarizes the electron spins, the first
term remains dominant for minority electrons. For majority electrons,
however, there is a range of times $\tau \ll \tau_{B}$ in
which the second term takes over, with
\begin{equation}
  \tau_{B} = \frac{a}{v} e^{2 E_{\rm Z}/g k_B T}.
\end{equation}
Hence, for times $\tau \ll \tau_{\rm B}$, $G^{>} \propto
\tau^{-1/2g - g/2}$, whereas $G^{>} \propto \tau^{-1/2 g}$ for $\tau
\gg \tau_{\rm B}$. Qualitatively, the crossover occurs when 
the electrons have a sufficiently long time $\tau_B$ to explore  
distances over which the electron spins are disordered. With 
increasing polarization that time increases. Note that the range
$\tau \ll \tau_{\rm B}$ is observable only if $\tau_{\rm B} \gg
a/v$, which requires polarizations $p_{\uparrow}$ close to unity.
Conversely, at any given $\tau$ (corresponding to a fixed bias 
voltage $V$) the tunneling current of majority spin electrons 
displays a strong magnetic field dependence: the dependence is
exponential $\exp(-E_{\rm Z}/k_B T)$ for $E^*_{\rm Z} \gtrsim E_{\rm Z} \gg k_B T$ and
saturates in fields larger than the crossover field $B^*$ with
\beq
E_Z^* = \frac{g}{2} k_B T \ln \frac{v\tau}{a}.
\eeq 

\subsection{Crossover as a function of distance from a boundary}
\label{length}

The anomalous scaling of spin-incoherent electrons can not only be
suppressed by polarizing the spins, but also by moving the tunneling
probe toward the end of the wire. Close to the end of the wire,
the sliding of the spin background is suppressed, and one expects to
recover the standard tunneling exponents for times $\tau \gg
\tau_x \equiv x/v$.
For $\tau \gg \tau_{\rm B}$ (when the conductor exhibits an 
anomalous exponent for tunneling into the bulk) we find
\begin{eqnarray} \label{Gx}
  G_\sigma^>(\tau)  &\propto &
  \frac{e^{i E_{\rm Z} \sigma \tau/2 }}{ w a}
  \frac{1-p_\sigma}{1+p_\sigma} 
   \\ && \mbox{} \times
   \left[\frac{ 4x^2 a^2  }{ \left((iv\tau+a)^2+4 x^2\right)(iv\tau+a)^2} \right]^{1/4g}  .\nonumber
\end{eqnarray}
Note that one indeed recovers the behavior $G^{>}(\tau) \propto \tau^{-1/g}$
for tunneling near the end of a standard Luttinger Liquid if $\tau
\gg \tau_x$.

In the regime $\tau_B \ll \tau_x$ one expects a double crossover 
for majority electrons: For very small times $\tau \ll \tau_{B}$,
one observes the exponent characteristic of a fully polarized electron
gas, $G^{>}(\tau) \propto \tau^{-1/2g-g/2}$. For the intermediate
range $\tau_{B} \ll \tau \ll \tau_x$, tunneling shows the anomalous
exponent characteristic of the bulk spin-incoherent Luttinger Liquid,
$G^{>}(\tau) \propto \tau^{-1/2 g}$. The true asymptotic behavior sets
in only for $\tau \gg \tau_x$ and shows the power-law dependence
$G^{>} \propto \tau^{-1/g}$ for to-end tunneling of a standard
Luttinger Liquid. Both crossovers can be inferred from the form of
the Green function at $\pi |w| \gg 1$ that is obtained by substituting
Eq.\ (\ref{capprox}) into Eq.\ (\ref{Gdir}) without taking any further 
limits,
\begin{widetext}
 \bea \label{Gcross}
G_\sigma^>(x,x,\tau) &\propto&  e^{i E_{\rm Z} \sigma \tau/2  }\left[\frac{ 4x^2 a^2}{ \left((iv\tau+a)^2+4 x^2\right)(iv\tau+a)^2} \right]^{1/4g} \left\{\frac{\sqrt{2}}{\sqrt{\pi}w}\,\frac{1-p_\sigma}{1+p_\sigma}  +\left[ \frac{a^2 \left((iv\tau+a)^2+4 x^2\right)(iv\tau+a)^2}{ 4x^2 }\right]^{g/4} \right. \nonumber \\
&&\mbox{}\left. \times \left[\frac{2}{(iv\tau+a)^g} +\frac{ 1}{ \left((iv\tau+a)^2+4 x^2\right)^{(g-1)/2 }}\left( \frac{e^{2 ik_F x}}{iv\tau-2ix+a}+ \frac{e^{-2 ik_F x}}{iv\tau+2ix+a}\right)\right]\right\} .
\eea
\end{widetext}
 While the first crossover at $\tau \approx \tau_B$ is obtained by comparing the magnitudes of the two terms inside the curly brackets, the second one at $\tau \approx \tau_x$     is described by the factor outside the curly brackets in Eq.\ (\ref{Gcross}).  

\section{Consequences for tunneling experiments}
\label{sec:4}
 
We now discuss the implications of Eqs.\ (\ref{Gh}), (\ref{Gx}), and
(\ref{Gcross}) for tunneling experiments. We consider two cases:
tunneling from one spin-incoherent wire into another one and
tunneling from a noninteracting conductor, such as a metallic STM-tip,
into a spin-incoherent wire.

 \begin{table}[htbp]
    \begin{tabular}{c|c|c|c} \hline
configuration: & end-end & bulk-end & bulk-bulk \\ \hline
 $\alpha_{\rm pol}$  & $2/g-1$ & $3/2g+g/2-1$ & $1/g+g -1$ \\  \hline
$\alpha_{\rm incoh}$  & $2/g-1$ & $3/2g -1$ & $1/g -1$  
   \\ \hline
\end{tabular} 
\caption{Scaling exponents for the tunneling currents between two
  spin-incoherent conductors subject to a magnetic field in various
  configurations. The spin-polarized exponents $\alpha_{\rm pol}$
  govern the  tunneling current at high voltages.  The crossovers
  between them and the incoherent exponents $\alpha_{\rm incoh}$ occur
  at the voltage scales $e V_B= E_F  \exp(-2E_{\rm Z}/gk_BT)$ for
  tunneling from bulk to end and $e V^*_B= E_F  \exp(-E_{\rm Z}/gk_BT)$ for bulk to bulk tunneling.} \label{table1} 
 \end{table}

In the first case, there is no real difference between `probe' and
`wire'. As a result, the exponent $\alpha$ governing the power-law
dependence of the tunneling current $I$ on the applied voltage $V$, 
\begin{equation}
  I \propto V^{\alpha},
\end{equation}
depends on whether the tunneling takes place from the bulk or from 
the end of the probe wire. 
The various exponents $\alpha_{\rm incoh}$ (characteristic of the
spin-incoherent Luttinger Liquid) and $\alpha_{\rm pol}$
(characteristic of the spin-polarized electron liquid) for tunneling
from end to end, bulk to end, and bulk to bulk respectively, are
summarized in table \ref{table1}. The crossovers between bulk and
end behavior occur at the corresponding scale $V \simeq V_x$. 

For tunneling from the end of the probe, the density of states in
the probe is independent of energy. Hence, the crossovers between 
various power law exponents for majority spin electrons described by  
Eqs.\  (\ref{Gh}) and (\ref{Gcross}) are observable at the scales 
derived in sections \ref{magnetic} and \ref{length}. The magnetic 
field dependence of the tunneling current in this configuration is 
given by that of the majority spin Green function derived in section
\ref{magnetic}.   

For tunneling from the bulk of the probe wire  into the bulk of the probed wire the
crossover takes place between a regime of high energies where the
majority spin current dominates to a low energy regime where the
current is carried by minority spin electrons. The fact that the minority spin
current dominates at low bias can be seen from the low voltage limit $I_\sigma \propto
p_{\sigma}(1-p_{\sigma})^2$ if $\tau \gg \tau_{B}$. It follows from the long-time asymptote
of Eq.\ (\ref{Gh}), used in Eq.\ (\ref{tunnel}) together with Eq.\ (\ref{Glesser}). The voltage at
which the minority and majority spin currents are equal is 
\begin{equation}
  e V_{B}^* \simeq E_F e^{-E_{\rm Z}/g k_B T}.
\end{equation}
At high voltages, $V \gg V_B^*$, the power law exponent $\alpha$ is that
characteristic of a fully polarized wire $\alpha_{\rm pol}$. At low
voltages, one observes the exponent of a spin-incoherent wire,
$\alpha_{\rm incoh}$, see Tab.\ \ref{table1}.
 
If tunneling takes place from a noninteracting conductor, the 
phase $\exp(i E_{\rm Z} \sigma\tau/2)$ in the Green function of the 
spin-incoherent wire becomes important. (This phase factor cancels
for tunneling between spin-incoherent conductors.) The phase
factor $\exp(i E_{\rm Z} \sigma\tau/2)$ represents the extra energy cost 
$E_{\rm Z}$ necessary to tunnel minority spin electrons
into the spin-incoherent wire. This energy cost appears because,
while the higher magnetic energy of the
minority electrons is compensated for by a reduction of their kinetic 
energy in the non-interacting probe, this is 
not possible in the spin-incoherent conductor, where 
the holes $c$ have a kinetic energy that is independent of the spin 
of the missing electron. As a consequence,  the chemical potentials of the spinless charge carriers in the
spin-incoherent conductor and of the electrons in the probe  at zero temperature differ 
 by the amount $E_{\rm Z}/2$, the energy gain $E_{\rm
  Z}/2$ for adding a majority spin to the wire.

We now assume that  $E_{\rm Z} \gg k_B T$, such that  $p_\downarrow \ll p_\uparrow$.
The current between a spin-incoherent conductor and a non-interacting probe  carried by
the majority electrons follows the scaling behavior reflecting the
time-dependence of the majority-spin Green function. The current
carried by the minority spins, however, consists of the `drift
current' at energy drop $E_{\rm Z} - e V$ 
for electrons exiting the 
spin-incoherent conductor and a `diffusion current' of thermally
activated electrons entering the spin-incoherent conductor from the 
probe. The drift and diffusion currents of minority electrons are
equal at zero bias, because the drift current of minority electrons 
exiting the spin-incoherent conductor is weighed with a factor
$p_{\downarrow} \simeq \exp(-E_{\rm Z}/k_B T)$, cf.\ Eq.\  (\ref{Glesser}). 

As a consequence the magnitude of the minority spin current  strongly depends on the direction of current flow.   For tunneling of electrons into the spin-incoherent wire the minority spin current dominates at voltages in the range $E_Z\ll eV \ll E_F$. This is a result of the factor $1-p_\sigma$ multiplying the first term in Eq.\ (\ref{Gh}) that is much larger for minority spins than for majority spins. For current flow in the opposite direction, however, the minority spin current is suppressed by the factor $p_\sigma$ in Eq.\ (\ref{Glesser}). The current at high voltages $V\gg V_B$  is then predominantly carried by majority spin electrons. The cross-over between the exponents 
$\alpha_{\rm pol}$ at $V \gg V_{B}$ and $\alpha_{\rm incoh}$ at $V \ll V_{B}$ derived from the form of the majority spin Green function in section \ref{magnetic} is observable if the crossover voltage $V_B = (E_F/e) \exp(-2
E_{\rm Z}/g k_B T)$  is larger than
$E_{\rm Z}/e$.    Also in this case of tunneling of electrons out of the spin-incoherent conductor the  current  is carried by minority electrons at sufficiently low voltages. This can be seen from the voltage dependence of the minority spin current  in the regime $eV \ll E_Z$. It resembles that of a semiconductor $p$-$n$ junction, 
$I_\downarrow \propto p_{\downarrow} E_F
(E_{\rm Z}/E_F)^{1/2 g} [1 - \exp(e V/k_B T)]$, and it  dominates over the majority spin current whose low voltage asymptote is $I_\uparrow \propto p_{\downarrow} E_F (eV/E_F)^{1/2 g}$.


\acknowledgments

This work was supported by the NSF
under grant no.\ DMR 0334499 and by the Packard Foundation.


\begin{thebibliography}{27}
\bibitem[{foo()}]{footnote}
 In this article we neglect logarithmic prefactors, assuming that they are dominated by power laws with nonzero exponents.
\end{thebibliography}
 \end{document}